\documentclass[a4paper,fleqn,usenatbib]{mnras}

\usepackage[T1]{fontenc}
\usepackage{ae,aecompl}

\usepackage{graphicx}	
\usepackage{amsmath}	
\usepackage{amssymb}	

\newcommand{\rev}{}

\title[A strong-lensing elliptical in MaNGA]
{A strong-lensing elliptical galaxy in the MaNGA survey}
\author[Russell J. Smith]{Russell J. Smith\thanks{E-mail: russell.smith@durham.ac.uk}
\\Center for Extragalactic Astronomy, University of Durham, Durham DH1 3LE\\
}

\date{Submitted to MNRAS, 5th August 2016; revised version 31st August.}

\pubyear{2016}

\begin{document}
\label{firstpage}
\pagerange{\pageref{firstpage}--\pageref{lastpage}}
\maketitle

\begin{abstract}
I report discovery of a new galaxy-scale gravitational lens system, identified using public data from  
the MaNGA survey, as part of a systematic search for lensed background line emitters. 
The lens is SDSS J170124.01+372258.0, 
a giant elliptical galaxy with velocity dispersion $\sigma$\,=\,256\,km\,s$^{-1}$, at a redshift of $z_{\rm l}$\,=\,0.122.
After modelling and subtracting the target galaxy light, the integral-field data-cube reveals 
[O\,{\sc ii}], [O\,{\sc iii}] and H$\beta$ emission lines corresponding to a source at $z_{\rm s}$\,=\,0.791,
forming an identifiable ring around the galaxy centre. 
{\rev If the ring is formed by a single lensed source, then} the Einstein radius is $R_{\rm Ein}$\,$\approx$\,2.3\,arcsec, 
projecting to $\sim$5\,kpc at the distance of the lens. 
The total projected lensing mass is  $M_{\rm Ein}$\,=\,(3.6$\pm$0.6)$\times$10$^{11}$\,M$_\odot$, and the total J-band
 mass-to-light ratio is 3.0$\pm$0.7 solar units. Plausible estimates of the likely dark matter content could reconcile this
 with a Milky-Way-like initial mass function 
 (for which $M/L$\,$\approx$\,1.5 is expected), but heavier IMFs are by no means excluded 
 with the present data.
{\rev An alternative interpretation of the system, with a more complex source plane, is also discussed.}
The discovery of this system bodes well for future lens searches based on MaNGA and other integral-field spectroscopic surveys. 
\end{abstract}



\section{Introduction}

Strong gravitational lensing provides the most precise and accurate means to determine absolute masses for distant galaxies
\citep{2010ARA&A..48...87T}. Such measurements  yield valuable constraints on the structure of dark-matter halos, and on
the normalisation of stellar mass estimates, independent of assumptions for the stellar initial mass function (IMF)
(e.g. \citet{2010ApJ...724..511A}).

Numerous methods have been developed to identify strong-lensing galaxies in a systematic fashion, e.g.
via their characteristic arc morphologies \citep{2006ApJ...638..703B}; from unusual colour configurations in catalogue data
\citep{2014ApJ...785..144G}; exploiting amplification above an underlying population e.g. for sub-mm galaxies \citep{2010Sci...330..800N},
or through  searches for multiple  components in the spectra of survey target galaxies
\citep{2006ApJ...638..703B}.

In the SNELLS (SINFONI Nearby Elliptical Lens Locator Survey) project, \citet*{2015MNRAS.449.3441S} adapted the spectroscopic method
to apply to integral-field unit (IFU) data. 
The specific aim of SNELLS was to discover strong-lensing galaxies at low-redshift, 
where (all other factors being equal) the Einstein radius is smaller, relative to the stellar effective radius, than in distant systems.
Lensing masses for such galaxies are subject to smaller fractional dark matter contributions, and so provide especially robust
limits on the stellar mass-to-light ratio, and hence the IMF \citep{2013MNRAS.434.1964S}.
However, since massive galaxies are intrinsically rare, the
probability that any nearby galaxy acts as a lens to a {\it bright} background source is very small.
To tackle this challenge, our method exploits the combined spatial and spectral contrast of IFU data to detect
{\it faint} background emission lines, and hence finds less spectacular, yet still scientifically valuable lens systems.
For SNELLS, we applied this method to targetted infra-red IFU observations of 27 very massive elliptical galaxies at $z_{\rm l}$\,$\la$\,0.06, 
discovering multiply-imaged background sources behind two of them, and recovering another previously known low-$z$ lens.

Despite these efforts, there are still very few  known lenses with $z_{\rm l}$\,$\la$\,0.1, and consequently the
low-redshift lensing approach to the IMF is uncomfortably dependent on the properties of the three galaxies discussed in 
\citet{2015MNRAS.449.3441S}. The advent of large galaxy surveys conducted using instruments with multiple deployable IFUs
promises to enlarge significantly the available sample of nearby strong lenses. 
The two leading surveys of this kind are SAMI (Sydney--Australian-Astronomical-Observatory Multi-object Integral-Field Spectrograph)
\citep{2015MNRAS.447.2857B} at the Anglo-Australian Telescope, and MaNGA  
(Mapping Galaxies at Apache Point Observatory) 
\citep{2015ApJ...798....7B}, which is part of the fourth-generation Sloan Digital Sky
Survey (SDSS). Both will gather IFU data for thousands of galaxies, with samples
that are weighted towards large, fairly low-$z$ galaxies, and hence suitable for finding stellar-mass-dominated lenses. 
The wide wavelength coverage (0.36--1.00\,$\mu$m) of MaNGA is especially advantageous for probing background emitters
over a large redshift range.

In this  {\it Letter}, I report a strong-lensing galaxy discovered from the first 
public MaNGA data \citep{2016arXiv160708619L}, from  SDSS Data Release 13 (DR13) (Albaretti et al. 2016, {\it in preparation}). 
Although this
particular system is probably too distant to provide very robust constraints on the IMF, it is a useful demonstration of the power of multiple-IFU
surveys to identify strong lenses which were (as in this case) not discovered in previous generation single-fibre data for the same 
galaxy. I briefly describe the data processing and search 
methods used in Section~\ref{sec:meth}, and the properties of the first newly-discovered lens system 
in Section~\ref{sec:lens}. Some preliminary analysis and comparison with other lenses is made in Section~\ref{sec:disc},
and summarized in Section~\ref{sec:concs}. A more thorough description of the MaNGA strong-lens search will be 
presented in a future paper, while detailed study of the system reported here will require follow-up observations.

\section{Search method}~\label{sec:meth}

For this preliminary study, I restricted the analysis to the 81 galaxies in MaNGA DR13 having 
SDSS {\rev single-fibre} velocity dispersion $\sigma$\,$>$\,250\,km\,s$^{-1}$. This is a small enough sample that all targets 
could be carefully inspected visually, and because the lensing cross-section scales  
approximately as $\sigma^4$, the total lensing probability for the survey is weighted heavily to these most massive 
galaxies\footnote{\rev{Galaxies with $\sigma$\,$>$\,250\,km\,s$^{-1}$ account for $\sim$40 per cent of $\sum\sigma^4$ for 
MaNGA DR13.}}.

As with the original SNELLS method, the broad approach is to extract an average spectrum for the primary target galaxy,  use this
as a simple model to remove the target light from each pixel in the data-cube, and then to examine suitably noise-normalised residual 
images by eye. 
For MaNGA, initial explorations showed that the combined data-cubes from DR13 are affected by small reconstruction 
artifacts,
{\rev causing abrupt changes in the spectrum shape on a pixel-by-pixel basis.
These features seem to arise from imperfect relative flux calibration, combined with masking of pixels affected by cosmic rays.
This leads to excessive spurious features in the residual data-cube, especially near the galaxy centres, which would
badly hamper detection of real emission lines.}

Thus, for this analysis, I instead work from the ``row-stacked spectra'' data products, which contain the individual input spectra 
and associated masks, {\rev in order to handle the  lens galaxy subtraction prior to any interpolation between fibres.}
%
%
%
%
The mean galaxy signal was subtracted from each spectrum
after modulation by a continuum fit to account for sensitivity and broad colour variations, and
the residuals in each wavelength channel were normalised by the median absolute deviation of all pixels in that channel.
Finally, I reconstructed the residuals to a new data-cube defined on a square-pixel oversampled grid. Specifically, the output pixels 
are 1\,arcsec, but each is an average of the residual flux over a 2$\times$2\,arcsec cell, hence neighbouring pixels are correlated.
The residual data-cube was then also output as an ``unwrapped cube'' (pseudo-long-slit) format, which is 
convenient for visual inspection.

In most cases the ``inspection images'' are fairly featureless. Residuals from strong sky lines are usually present, but largely supressed by
the noise normalisation. Where target galaxies have spatially-varying spectra, e.g. strong emission lines, or evident rotation, 
the effects are visible in the inspection images, but easily distinguished from the localised background emission signal being sought here. 

Background line-emitters are detected in the inspection images for around a quarter of all targets considered. In almost all cases,
multiple lines are visible to confirm the source redshift, {\rev while for one distant [O\,{\sc ii}] emitter the identification was
clear based on the resolved doublet structure.}
From the DR13 $\sigma$\,$>$\,250\,km\,s$^{-1}$ sample, all but one of the background emitters were located at large $\ga$5\,arcsec from
the primary target, and hence unlikely to be multiply imaged.
Curiously, several targets have two background galaxies at a
common or very similar redshift, but the separation is so large in most cases that they must be multiple sources, e.g. in a background group.

Only one system was discovered among this sample that can be confidently classified as lensed, based on the MaNGA data alone, 
as described in the next section.

\section{A bright Einstein ring in J1701+3722}~\label{sec:lens}

\begin{figure}
\includegraphics[width=85mm]{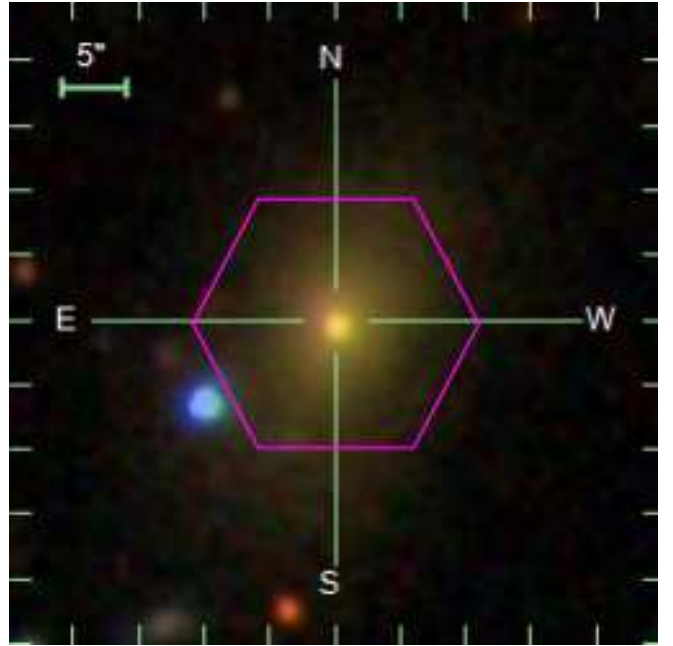}
\vskip 0mm
\caption{The MaNGA lens galaxy J1701+3722. The lensed source is not visible in this SDSS broad-band image, but the
faint feature just inside the south-east corner of the MaNGA field is the ``southern source'' at the same redshift as the lensed galaxy.
{\rev The bright blue object outside the IFU is a foreground star.}}
\label{fig:colimg}
\end{figure}

\begin{figure*}
\includegraphics[width=180mm]{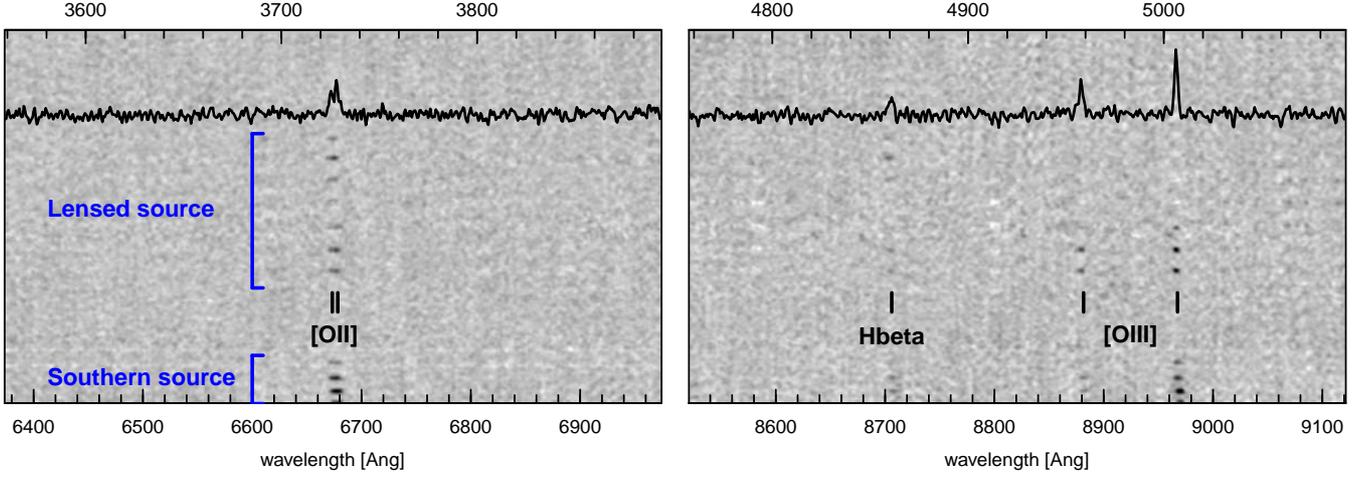}
\vskip 0mm
\caption{Extracts from the ``inspection image'' for the MaNGA lens galaxy J1701+3722. This is an
unwrapped cube or pseudo-long-slit spectrum, after removing the $z$\,=\,0.122 target galaxy signal, and normalising by the
noise spectrum. Positive signals (emission lines) are dark in this image. 
The vertical direction follows the spatial raster, with the north of the field at the top. 
{\rev The wavelength scales are as observed (lower) and at rest for $z$\,=\,0.791 (above). The trace shows
the noise-normalised residual spectrum extracted from the brightest part of the lensed source.}
}
\label{fig:inspec}
\end{figure*}

\begin{figure*}
\includegraphics[width=125mm]{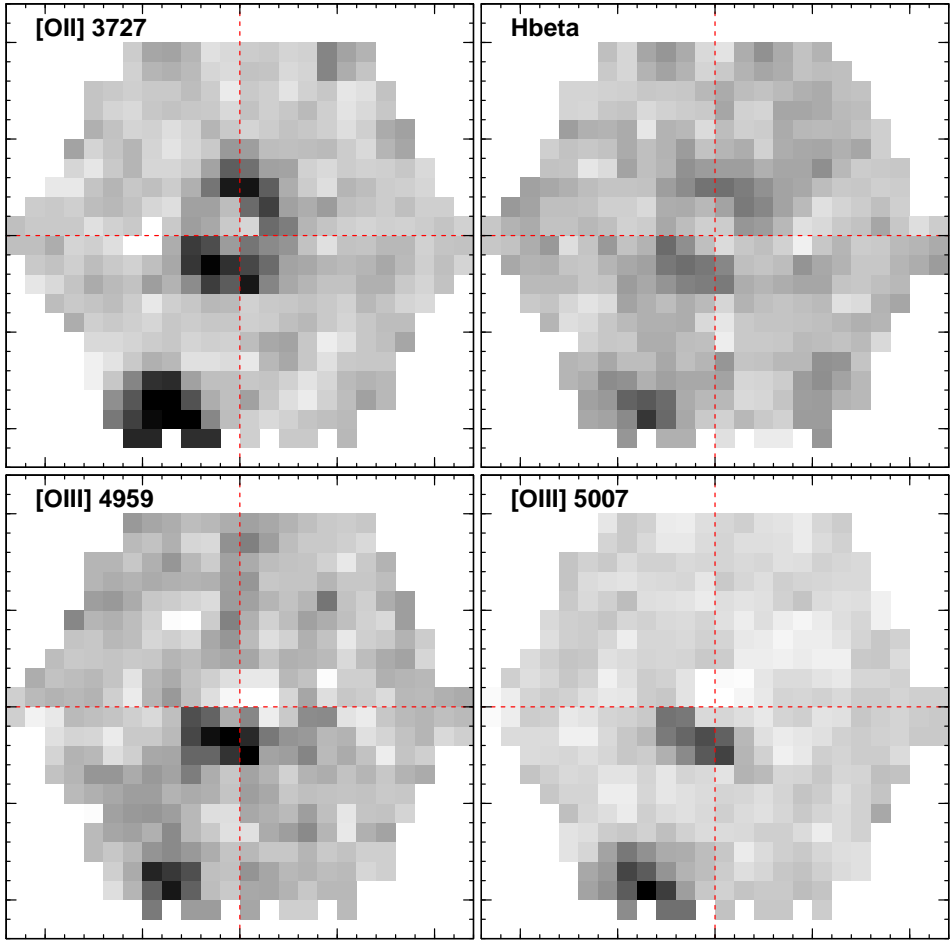}
\vskip 0mm
\caption{Spatial structure of the four detected $z$\,=\,0.791 background emission lines in  J1701+3722. The panels are 24\,arcsec on each side, 
with 1\,arcsec pixels, with the same orientation as Figure~\ref{fig:colimg}.
The colour-scaling is slightly adjusted in each panel to emphasize the main features. 
An apparent Einstein ring configuration is seen in the upper panels, but only one ``arc'' is bright in [O\,{\sc iii}].
 The ``southern source'' at the same redshift
is seen in the corner of the field.
}
\label{fig:ring}
\end{figure*}

A strongly-lensed background line-emitter was discovered in the residual data-cube for 
target SDSS J170124.01+372258.0 (hereafter J1701+3722)\footnote{\rev{Also identified as
MaNGA ID 1-136292, and MaNGA plate--ifu number 8606-6102.}}, which is a giant elliptical galaxy with 
$\sigma$\,=\,256$\pm$8\,km\,s$^{-1}$ and redshift $z$\,=\,0.122. 
The SDSS image and MaNGA IFU field for this target are shown in Figure~\ref{fig:colimg}.
The galaxy image appears smooth and without any peculiarities.
The original SDSS single-fibre spectrum exhibits a pure continuum spectrum,
characteristic of an old stellar population, with no emission lines to indicate ongoing star formation or nuclear activity.

Figure~\ref{fig:inspec} shows an extract from the MaNGA inspection image for this galaxy, 
which reveals prominent emission lines that are consistent with
[O\,{\sc ii}] 3727\,\AA, H$\beta$,  [O\,{\sc iii}] {\rev 4959}\,\AA\ and [O\,{\sc iii}] 5007\,\AA,
at a redshift of $z$\,=\,0.791. As with some other objects found in this project (as mentioned above), 
the emission is clearly broken into several spatial parts. 
The upper section itself displays two bright regions separated by a fainter section. The uppermost part is clear only in  [O\,{\sc ii}], but also
faintly visible in H$\beta$. The  [O\,{\sc ii}] doublet line profile is identifiable throughout both sources; the redshift is unambiguous.

Figure~\ref{fig:ring} shows the spatial structure of the $z$\,=\,0.791 emission, from the four brightest lines. 
The brightest part of the emission corresponds to a separate object at the same redshift, 
well to the south of the target galaxy. This source is faintly visible in the SDSS imaging. 
In the  [O\,{\sc ii}] line, the northern part of the emission seems to form a near-complete ring, with radius 
$\sim$\,2.3\,arcsec around the centre of J1701+3722. 
The ring is possibly broken into two ``arcs'', though the detailed morphology should not be over-interpreted at this stage, given
the low resolution of the reconstructed image, and the processing steps applied.
{\rev Within the ring,  [O\,{\sc iii}] 5007\,\AA\  is the brightest line, peaking at $\sim$1.5$\times$10$^{-16}$\,erg\,s$^{-1}$\,cm$^{-2}$ per MaNGA fibre,
while  [O\,{\sc iii}] 3727\,\AA\ reaches  $\sim$1.0$\times$10$^{-16}$\,erg\,s$^{-1}$\,cm$^{-2}$ per MaNGA fibre.}
Curiously, the emission morphology appears to be  different in the [O\,{\sc iii}] lines, where only the southern ``arc'' is visible,
both ``arcs'' are seen in the weak H$\beta$ line.

\section{Discussion}~\label{sec:disc}

The newly-discovered lens occupies an interesting redshift niche, being substantially more distant than the 
SNELLS galaxies (at $z$\,=\,0.03--0.05), but at near end of the range probed by the
SLACS (Sloan Lens Advanced Camera for Surveys) sample of \citet{2008ApJ...682..964B} 
($z$\,=\,0.10--0.35). The velocity dispersion is smaller than that of the SNELLS lenses, but
higher than average for the closest lenses in SLACS. The background source in 
J1701+3722 is so bright that it may be surprising it wasn't detected from the 
single-fibre SDSS spectroscopy: the  [O\,{\sc ii}] line can, with hindsight, be clearly seen as a residual peak
in the original spectrum, but the other lines cannot. The  [O\,{\sc ii}]  and  [O\,{\sc iii}] lines
{\it are} both seen, however, in a 3-arcsec ``aperture'' constructed from the deeper  observation for MaNGA.

{\rev Assuming the simplest interpretation for the system, i.e. an Einstein ring formed by a single source (but see below), 
the total projected mass can be estimated from the Einstein radius, using the symmetric lens approximation.}

Adopting $R_{\rm Ein}$\,=\,2.3\,$\pm$0.2\,arcsec, together with the angular diameter
distances  computed with the same \citet{2011ApJS..192...18K} cosmological parameters as used in \citet{2015MNRAS.449.3441S},
this yields  $M_{\rm Ein}$\,=\,(3.6$\pm$0.6)$\times$10$^{11}$\,M${_\odot}$. Again following \citet{2015MNRAS.449.3441S} for consistency, 
I estimate the luminosity in the J-band, from 2MASS (the 2-Micron All-Sky Survey, \citep{2006AJ....131.1163S}).
J1701+3722 has $J$\,=\,14.8$\pm$0.2 in the  2.3-arcsec aperture, corresponding
to $L_{\rm Ein}$\,=\,(1.2$\pm$0.2)$\times$10$^{11}$\,L${_\odot}$. Hence the total mass-to-light ratio inside the Einstein aperture is
3.0$\pm$0.7. For comparison the total mass-to-light ratio of the three SNELLS lenses was 1.8--2.1 in the same band.
According to the \cite{2005MNRAS.362..799M} models,  the expected value for old stellar populations alone (i.e. no dark matter) 
is $\sim$1.5 with a \cite{2001MNRAS.322..231K} IMF, or $\sim$2.3 with a \cite{1955ApJ...121..161S} IMF.

The contribution of dark matter to the lensing mass in J1701+3722 will be larger than in the SNELLS galaxies, 
because the lens is more distant, so the Einstein radius projects to a larger physical scale: $\sim$5.0\,kpc,
compared to $\sim$2\,kpc for SNELLS. The method applied in \citet{2015MNRAS.449.3441S},
based on projected dark-matter halo profiles in simulations \citep{2015MNRAS.451.1247S}, 
yields an estimate of $M_{\rm DM}$\,=\,(1.2$\pm$0.2)$\times$10$^{11}$\,M${_\odot}$,  within $R_{\rm Ein}$,
i.e. $\sim$33 per cent of the total.
Thus the total mass-to-light ratio obtained here is consistent with the SNELLS results favouring a Milky-Way-like IMF, but given
the large present uncertainties, they are also quite compatible with Salpeter or heavier IMFs.


{\rev The estimates above should be treated with caution until the lens can be better characterized with improved data. 
Two factors in particular suggest that the simple lens model might be inappropriate for J1701+3722. Firstly, the 
radius of the observed ``ring'' is 50 per cent larger than the expected Einstein radius for an isothermal mass distribution with 
$\sigma$\,=\,256\,km\,s$^{-1}$, given the lens and source redshifts. Secondly, there is the different appearance of the system
in the  [O\,{\sc iii}]  versus the [O\,{\sc ii}] lines. This disparity is  probably physical,
rather than caused by artifacts in the observations or data processing, since both [O\,{\sc iii}] lines are affected similarly,
while H$\beta$ follows the same structure as [O\,{\sc ii}]. 
%
%
An alternative interpretation would be that the arcs correspond to two {\it different} sources at the same redshift, each 
being either singly-imaged or having a faint counter-image at smaller radius which is not detectable in the present data. 
This scenario would imply a smaller Einstein radius, resolving the apparent discrepancy with the measured velocity 
dispersion\footnote{I am grateful to Tom Collett for proposing this interpretation.}.

Future observations at higher spatial resolution are essential to distinguish between these possibilities.}

\section{Summary and outlook}~\label{sec:concs}

In this {\it Letter} I have presented the discovery of a strongly-lensed emission-line galaxy behind a $z$\,=\,0.122 elliptical
observed in the MaNGA survey. The survey data allow a crude initial estimate of the Einstein radius and
total lensing mass, {\rev under the simplest assumptions}, 
but improved data will be necessary to characterize the lensing configuration properly. 

The discovery of such a bright example, from just the first few dozen MaNGA observations analysed,
augurs well for identifying numerous strong lenses with the new generation of multi-IFU surveys.
I am currently extending this initial study to search for lenses among MaNGA galaxies with lower 
velocity dispersion, and also
conducting an equivalent programme using data from the SAMI survey.
While the method can certainly detect much fainter background sources than that in J1701+3722, most such systems will require
follow-up observations to confirm the candidates. Nonetheless, this approach is a promising means to 
enlarge the sample of low-redshift, stellar-mass-dominated lenses.

\section*{Acknowledgements} 
This work was supported by the STFC through grant ST/L00075X/1. {\rev The data used here are publicly
available from the SDSS science archive server (Data Release 13).}

{\rev I am grateful to Tom Collett for suggesting the ``two-source'' lensing interpretation, 
and to John Lucey for discussions about the work. David Law, as referee, provided 
prompt and helpful comments.} 

Funding for the Sloan Digital Sky Survey IV has been provided by
the Alfred P. Sloan Foundation, the U.S. Department of Energy Office of
Science, and the Participating Institutions. SDSS-IV acknowledges
support and resources from the Center for High-Performance Computing at
the University of Utah. The SDSS web site is www.sdss.org.

SDSS-IV is managed by the Astrophysical Research Consortium for the 
Participating Institutions of the SDSS Collaboration including the 
Brazilian Participation Group, the Carnegie Institution for Science, 
Carnegie Mellon University, the Chilean Participation Group, the French Participation Group, Harvard-Smithsonian Center for Astrophysics, 
Instituto de Astrof\'isica de Canarias, The Johns Hopkins University, 
Kavli Institute for the Physics and Mathematics of the Universe (IPMU) / 
University of Tokyo, Lawrence Berkeley National Laboratory, 
Leibniz Institut f\"ur Astrophysik Potsdam (AIP),  
Max-Planck-Institut f\"ur Astronomie (MPIA Heidelberg), 
Max-Planck-Institut f\"ur Astrophysik (MPA Garching), 
Max-Planck-Institut f\"ur Extraterrestrische Physik (MPE), 
National Astronomical Observatory of China, New Mexico State University, 
New York University, University of Notre Dame, 
Observat\'ario Nacional / MCTI, The Ohio State University, 
Pennsylvania State University, Shanghai Astronomical Observatory, 
United Kingdom Participation Group,
Universidad Nacional Aut\'onoma de M\'exico, University of Arizona, 
University of Colorado Boulder, University of Oxford, University of Portsmouth, 
University of Utah, University of Virginia, University of Washington, University of Wisconsin, 
Vanderbilt University, and Yale University.

\bibliographystyle{mnras}
\bibliography{rjs} 

\bsp	
\label{lastpage}
\end{document}